\documentclass[usegraphicx]{mn2e}
\usepackage{times}
\usepackage{graphicx,subfigure}

\begin{document}

\title[Foreground contribution of dust towards Kepler]
      {Accounting for the Foreground Contribution to the Dust Emission towards Kepler's Supernova Remnant.}

      \author[H.\ Gomez et al.]  {H.\,L.\ Gomez,$^{\!
          1}$\thanks{E-mail: haley.gomez@astro.cf.ac.uk} L.\
        Dunne,$^{\! 2}$ R.\,J.\ Ivison,$^{3,4}$ E.\,M.\ Reynoso,$^{\!
          5,6}$\thanks{Member of the Carrera del Investigador
          Cient\'{\i}fico of CONICET, Argentina}, M.\,A.\
        Thompson,$^{\! 7}$ \and B.\ Sibthorpe,$^{\!  3}$ S.\,A.\
        Eales,$^{\! 1}$ T.\,M.\ DeLaney,$^{\!8}$ S.\ Maddox,$^{\! 2}$
       and  K.\ Isaak$^{1}$ \\
        $^1$ School of Physics \& Astronomy, University of Wales,
        Cardiff
        CF24 3YB, UK\\
        $^2$ School of Physics \& Astronomy, University of Nottingham,
        University Park, Nottingham NG7 2RD, UK\\
        $^3$ UK Astronomy Technology Centre, Royal Observatory
        Edinburgh, Blackford Hill, Edinburgh EH9 3HJ, UK\\
        $^4$ Institute for Astronomy, University of Edinburgh,
        Blackford Hill, Edinburgh EH9 3HJ, UK\\
        $^5$ Instituto de Astronomia y F\'{\i}sica del Espacio (IAFE),
        CC 67, Suc. 28, 1428 Buenos Aires, Argentina\\
        $^6$ Departamento de F\'{\i}sica, Facultad de Ciencias Exactas
        y Naturales, Universidad de Buenos Aires, Argentina\\
        $^7$ Centre for Astrophysics Research, Science and Technology
        Research Institute, University of Hertfordshire, College Lane,
        Hatfield, AL10
        9AB, UK\\
        $^8$ MIT Kavli Institute, 77 Massachusetts Avenue, Room
        NE80-6079, Cambridge, MA 02139, USA}

\date{}

\pagerange{\pageref{firstpage}--\pageref{lastpage}} \pubyear{2009}

\maketitle

\begin{abstract}
  Whether or not supernovae contribute significantly to the overall
  dust budget is a controversial subject.  Submillimetre (submm)
  observations, sensitive to cold dust, have shown an excess at 450
  and 850\,$\mu$m in young remnants Cassiopeia~A (Cas~A) and
  Kepler. Some of the submm emission from Cas~A has been shown to be
  contaminated by unrelated material along the line of sight. In this
  paper we explore the emission from material towards Kepler using
  submm continuum imaging and spectroscopic observations of atomic and
  molecular gas, via H\,{\sc i}, $^{12}$CO ($J$=2--1) and $^{13}$CO
  ($J$=2--1).  We detect weak CO emission (peak $T_{\rm A}^*$ =
  0.2--1\,{\sc k}, 1--2\,km\,s$^{-1}$ {\sc fwhm}) from diffuse,
  optically thin gas at the locations of some of the submm clumps. The
  contribution to the submm emission from foreground molecular and
  atomic clouds is negligible.  The revised dust mass for Kepler's
  remnant is 0.1--1.2\,M$_{\odot}$, about half of the quoted values in
  the original study by Morgan et al.\ (2003), but still sufficient to
  explain the origin of dust at high redshifts.
\end{abstract}


\section{Introduction}

The conditions following a supernova (SN) explosion are thought to be
conducive to the formation of dust: the abundances of heavy elements
are high, as is the density; temperatures drop rapidly in the
expanding ejecta, quickly reaching levels allowing the sublimation of
grain materials. Theoretical estimates predict that type-{\sc ii} SNe
should produce a significant quantity of dust, approximately $\rm
0.3-5 \, M_{\odot}$ per star, depending on the metallicity, stellar
mass and energy of the explosion (e.g. Todini \& Ferrara 2001; Nozawa
et al.\ 2003; Schneider, Ferrara \& Salvaterra 2004). Some
circumstantial evidence also leads us to believe that SNe should be an
important source of dust: first, without SNe there is a dust budget
crisis in the Galactic ISM.  The dust produced in cool stellar
atmospheres of intermediate-mass stars, combined with current
predictions for how much dust is destroyed in shocks, yields far less
dust than is observed (Jones et al.\ 1994) in the interstellar medium
(ISM). Either another source of dust exists, or dust destruction
cannot be as efficient as is widely believed. Second, without SNe and
their massive precursors as significant sources of dust, it is
difficult to explain the immense dust masses found in
submillimetre-selected galaxies and quasars at high redshift (e.g.\
Smail et al.\ 1997; Isaak et al.\ 2002; Eales et al.\ 2003). There is
not sufficient time for dust to form in such large quantities from
evolved stars alone (Morgan \& Edmunds 2003; Dwek et al.\ 2007 and
references therein).

The signature of warm, freshly-formed dust in Cas~A was seen in
spectroscopic data by Rho et al.\ (2008), with reported dust masses in
the range 0.02--0.05-M$_{\odot}$. Smaller fractions of warm dust have
been reported in SN2003gd and Kepler (Sugerman et al.\ 2006; cf.\
Meikle et al.\ 2007; Blair et al. 2007). {\em Spitzer} observations of
SNRs in the Magellanic Clouds are also consistent with small amounts
of dust (e.g. Borkowski et al.\ 2006; Williams et al.\ 2006). However,
{\it Spitzer} is not sensitive to the presence of very cold dust,
which peaks at wavelengths longer than 160\,$\mu$m.  To address the
question of whether large quantities of dust are present, we require
observations at longer wavelengths in the submm.

The Submillimetre Common User Bolometer Array (SCUBA -- Holland et
al.\ 1999) on the James Clerk Maxwell Telescope (JCMT\footnote{The
  JCMT is operated by the Joint Astronomy Centre on behalf of the UK's
  Science and Technology Facilities Council, the Netherlands
  Organisation for Scientific Research, and the National Research
  Council of Canada.}) was used to observe the young Galactic SN
remnant (SNR) Cas~A (Dunne et al.\ 2003 -- hereafter D03), with large
excesses of submm emission detected over and above the extrapolated
synchrotron components. This was confirmed by ARCHEOPS (D\'{e}sert et
al.\ 2008). Because of the high spatial correlation with the X-ray and
radio emission, this was interpreted as emission from cold dust
associated with the remnant. However, some of the submm emission comes
from molecular clouds along the line of sight (Krause et al.\ 2004;
Wilson \& Batrla 2005).  The high degree of submm polarisation
suggests that a significant fraction of dust does originate within the
remnant ($\sim \rm 1\,M_{\odot}$, Dunne et al.\ 2009).  We now turn to
the only other remnant with a reported excess of submm emission over
the extrapolated synchrotron: Kepler.  We originally interpreted our
SCUBA data as evidence for 0.3--3\,$\rm{M_{\odot}}$ of dust associated
with the remnant (Morgan et al.\ 2003 -- hereafter M03; Gomez et al.\
2007), an order of magnitude higher than those predicted from 160-$\mu$m
{\em Spitzer} data of Kepler (Blair et al.\ 2007).

Kepler's SN has a shell-like structure, $\sim$3\,arcmin in
diameter. Estimates of its distance, using H\,{\sc i} absorption
features range from 3.9 to 6\,kpc (Reynoso \& Goss 1999 -- hereafter
RG99; Sankrit et al.\ 2008). Its classification has been controversial
(Blair et al.\ 2007; Reynolds et al.\ 2007; Sankrit et
al.\ 2008) with evidence pointing towards either a type-{\sc i}a --
the thermonuclear explosion of a low-mass accreting star in a binary
system -- or a type-{\sc i}b -- the core collapse of a massive star.
Reynolds et al.\ proposed that Kepler's SN was the result of a
thermonuclear explosion in a single 8-M$_{\odot}$ star, after roughly
50\,Myr of evolution, which would make it pertinent to the issue of
the dust budget in the early Universe. Of course, understanding dust
formation in SNRs is important regardless of the explosion mechanism:
dust formation following type-{\sc i}a SNe would indicate that
type-{\sc ii} SNe would also be likely dust producers (Clayton et al.\
1997; Travaglio et al.\ 1999).  Here, we revisit the submm data and
ask how much of the submm emission can be associated with the remnant
and how much with material along the line of sight.

In this paper, we recalculate the contribution of synchrotron
radiation to the submm emission in Kepler using a radio spectral index
map (kindly provided by T.\ DeLaney). We also present the first
high-resolution molecular-line map towards Kepler's SNR and compare
this with archival H\,{\sc i} data (RG99). In \S\ref{sec:submm} we present
details of the observations and data reduction; in \S\ref{sec:coobs}
we investigate the possibility of contamination by line-of-sight gas
clouds.  In \S\ref{sec:dustmass} we compare the distribution of the dust
and gas.  In \S\ref{sec:dust} we estimate the dust mass in Kepler with
results summarised in \S\ref{sec:conc}.  Simulations of the effects of
the SCUBA chop throw are discussed in Appendix\ref{sec:bruce}.

\section{Observations and analysis}
\label{oanda}

\subsection{Submillimetre continuum observations}
\label{sec:submm}

The reduction of SCUBA data of Kepler's remnant at 450 and
850\,$\mu$m, using standard routines in the SURF software package
(Sandell et al.\ 2001), was described briefly by M03. The observations
were carried out over five different nights during 2001--03 using the
jiggle-map mode.  The array was chopped to remove sky emission; we
chose reference positions using a map of the radio emission as a
guide, with a chop throw of 180\,arcsec (see Fig.~\ref{fig:bruce}).  The
potential hazards of this observing mode are discussed and simulated
in \S\ref{sec:bruce}.  In M03 we presented the synchrotron-subtracted
850-$\mu$m image, obtained by scaling the 5-GHz Very Large Array
(VLA\footnote{The VLA is operated by the National Radio Astronomy
  Observatory, which is a facility of the National Science Foundation,
  operated under cooperative agreement by Associated Universities,
  Inc.}) map of Kepler using a constant spectral index,
$S_{\nu}\propto \nu^{-0.71}$, that being the mean value reported by
DeLaney et al.\ (2002). We take a different approach here, instead
using the {\em spectral index} map of DeLaney et al.\ -- their
Figure~4 -- to produce a more accurate estimate of the synchrotron
contribution pixel-by-pixel.  The flatter spectral index in the north
of the remnant means that more flux has been subtracted from the submm
images than in the analysis of M03.  The final synchrotron-subtracted
submm flux densities are $0.4\pm 0.1$ and $2.8\pm 0.7$\,Jy at 850 and
450\,$\mu$m, respectively.

\begin{figure*}
\includegraphics[angle=0,width=17.5cm,height=6.5cm]{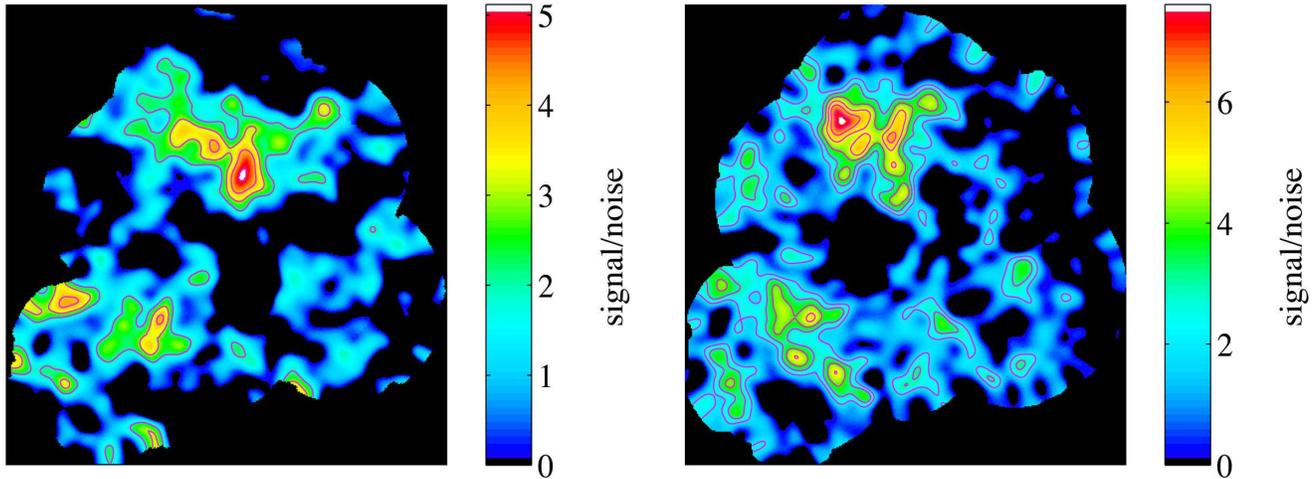}
\caption{Signal-to-noise submillimetre images of Kepler's SNR with synchrotron
  contribution removed. {\em Left}: 450\,$\mu$m with resolution
  16\,arcsec and signal-to-noise contours at 2, 3, 4, 5$\sigma$ ($\sigma
  \sim  46$\,mJy\,beam$^{-1}$).  {\em Right}: 850\,$\mu$m with resolution 19\,arcsec with contours 3, 4, 5, 6$\sigma$ ($\sigma \sim 8$\,mJy\,beam$^{-1}$).  The centre of the remnant is at $\rm \alpha_{2000}=17^h 30^m41^s.3$, $\rm \delta_{2000}= -21^{\circ} 29'29''$.}
\label{fig:keplersubmm}
\end{figure*}

The signal-to-noise maps of {\it cold} dust at 450 and 850\,$\mu$m are
shown in Fig.~\ref{fig:keplersubmm}, with the synchrotron contribution
subtracted using the spectral index image.  The noise images were
created by randomly generating 1,000 artificial images (Eales et al.\
2000). The noise is substantially higher near the edges of the map
because these regions received significantly less integration time. At
450\,$\mu$m, the centre of the map is also noisy (the array footprint
is smaller at 450\,$\mu$m than at 850\,$\mu$m). The peak
signal-to-noise values in these maps are 5$\sigma$ at 450\,$\mu$m and
7$\sigma$ at 850\,$\mu$m.  There are regions of high flux near the
edges of the map (particularly in the south-east) but these are also
regions of high noise so our confidence in these features is
low. Conversly, there are regions of low flux which are not
particularly noisy (e.g.\ in the southern `ring' region of the
remnant); our confidence in these features is commensurately higher.

\begin{figure}
\includegraphics[width=7.5cm,angle=-90]{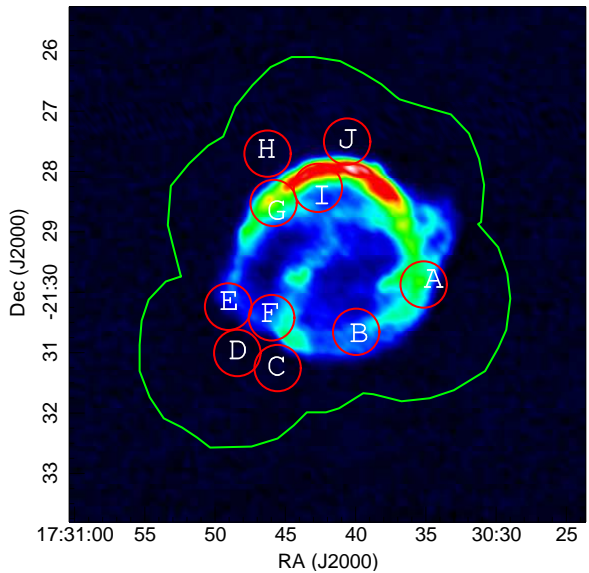}
\caption{1.4-GHz VLA image of Kepler's SNR. The rms noise
    is 0.6\,mJy\,beam$^{-1}$. Submm clumps are labelled A--J with radius,
    23\,arcsec. The green contour indicates the boundary of
    the SCUBA observations and the black box indicates the extent of
    the molecular observations in this work (\S~\ref{sec:coobs}).}
  \label{fig:clumps}
\end{figure}

The location of the submm peaks A--J, are marked on the 1.4-GHz map in
Fig.~\ref{fig:clumps}. This image was constructed using data from the
VLA archive (see \S\ref{sec:hi}). Dust clumps A, B, F, I and G all
fall within the radius of the shockfront. Most of the emission from
cloud E is within the shock also whereas C, D, H, and J lie beyond the
X-ray and radio boundaries. Cloud E is located at the position of one
of the `ears' in the radio, a region associated with the ejecta but
beyond the almost circular shock front. Around 30--40 per cent of the
submm flux lies outside the shock front (as defined by the radio
observations at 100\,arcsec). 

Possible explanations for the submm emission are: (i) dust produced by
the supernova remnant or progenitor star; (ii) interstellar material
along the line of sight and/or (iii) spurious structure in the SCUBA
map -- perhaps an artefact of the observing or data processing
techniques. These are possibilities we will explore in the remainder
of this paper.

\subsection{Exploring the possibility of line-of-sight contamination}
\label{sec:coobs}

\begin{figure*}
\includegraphics[angle=0,width=9cm]{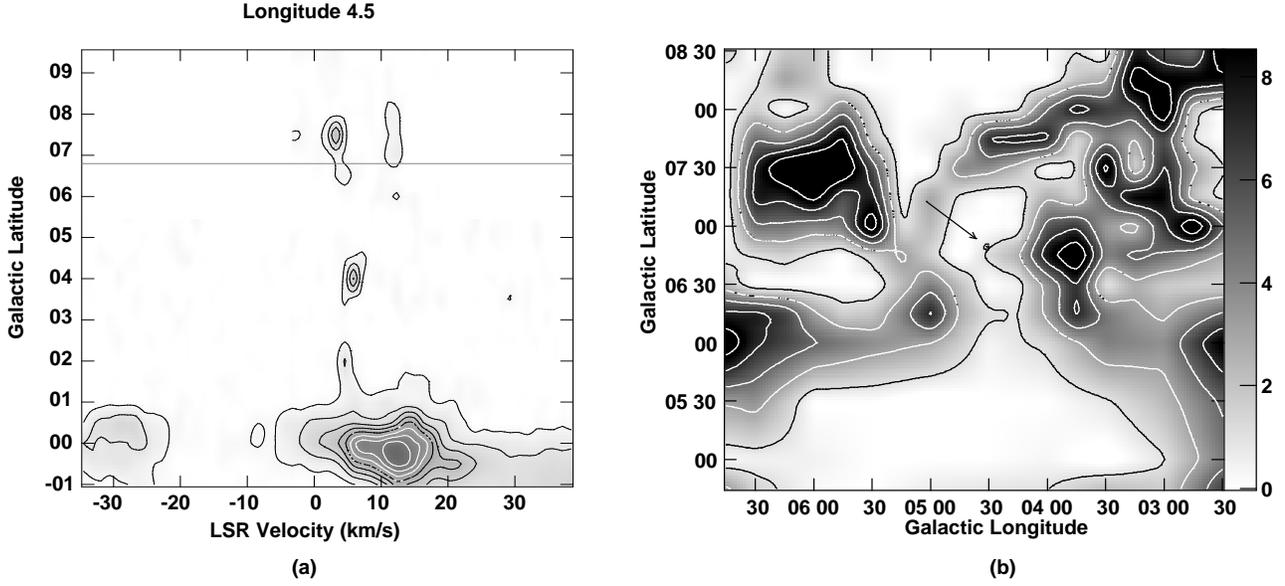}
\caption{(a) Latitude-velocity map at a longitude of
    4.5$^{\circ}$ from the CO Milky Way Survey (e.g.\ Dame et al.\
    2001). Contours show $T_{\rm A}^*$ ranging from 0.3 to 4.5\,{\sc
    k} in steps of 0.6\,{\sc k}. A line at a constant latitude of
    6.8$^{\circ}$ is included to show the location of Kepler's
    SNR. The greyscale goes from 0 to 8.5\,{\sc k}, as shown at the
    right of Fig.~3b. (b) Integrated CO image of the molecular clouds
    (part of the Ophiuchus complex at 165\,pc) in the vicinity of
    Kepler's SNR, taken from the Galactic CO data available at {\small \tt
    http://cfa-www.harvard.edu/mmw/MilkyWaynMolClouds.html}. The
    resolution is 8\,arcmin and the integration spans over $-300 < v <
    300$\,km\,s$^{-1}$. The contours are at constant $T_{\rm A}^*$,
    ranging from 1 to 9\,{\sc k} in steps of 1\,{\sc k}. The greyscale
    is indicated to the right of the image. An outer radio continuum
    contour of Kepler's SNR with location indicated by the arrow. }
\label{fig:dame}  
\end{figure*}

Does Kepler's SNR have significant interstellar material along the
line of sight that may be contributing to the measured submm fluxes?
Kepler's SNR is approximately 600\,pc out of the Galactic Plane ($l
\sim\rm 4.5^{\circ}$, $b \sim\rm 6.8^{\circ}$) with extinction and
100-$\mu$m background lower than measured at Cas~A by a factor of
three ({\it IRAS} IRSKY maps -- Arendt 1989).  A latitude-velocity map
of integrated $^{12}$CO ($J$=1--0) (kindly provided by T.\ Dame) at
the location of Kepler shows that the velocity range of clouds
integrated over longitude is confined to $-5< v < 5$\,km\,s$^{-1}$
(Fig.~\ref{fig:dame}a). Fig.~\ref{fig:dame}b shows the integrated CO
emission near Kepler's SNR (marked by the radio contour in the centre)
from the processed Galactic CO survey (Dame et al.\ 2001).  The remnant
is within 3$^{\circ}$ of the $\rho$~Oph cloud complex (distance,
165\,pc) which has clouds with line widths ($\Delta v$) up to
100\,km\,s$^{-1}$.

\subsubsection{$^{12}$CO($J$=2--1) observations: searching for molecular structures}
\label{sec:co}
In order to quantify the possible contribution from
foreground molecular material, we observed Kepler's SNR in the
$^{12}$CO($J$=2--1) line using the A3 receiver on the JCMT with the
Auto-Correlation Spectrometer Imaging System (ACSIS) in
double-sideband mode (DSB) (Table~\ref{tab:obs}).  The bandwith was
1.8\,GHz with 1,904 channels and a velocity coverage of
2,000\,km\,s$^{-1}$.  We raster-mapped over a region $\rm 8 \times
8$\,arcmin$^2$ with grid points separated by 7\,arcsec to avoid
smearing in the scan direction (the beam at this wavelength is
20.8\,arcsec {\sc fwhm}). The integration time was 5\,s at each point.
In this intial dataset, the lines were unresolved.  Further
high-resolution service observations were obtained to resolve the
lines at the location of two CO peaks.  The high-resolution data were
taken with the same receiver set-up, but with a bandwidth of 250\,MHz,
8,192 channels and a velocity coverage of 300\,km\,s$^{-1}$. Spectral
resolutions and positions are given in Table~\ref{tab:obs}.  The
spectra were baseline-subtracted with a third-order polynomial and
scaled to main beam ($T_{\rm MB}$) temperatures by dividing the
antenna temperatures ($T_{\rm A}^*$) by the telescope efficiency
$\eta_{\rm MB}\sim 0.69$.
\begin{figure*}
\includegraphics[angle=-90,width=18cm]{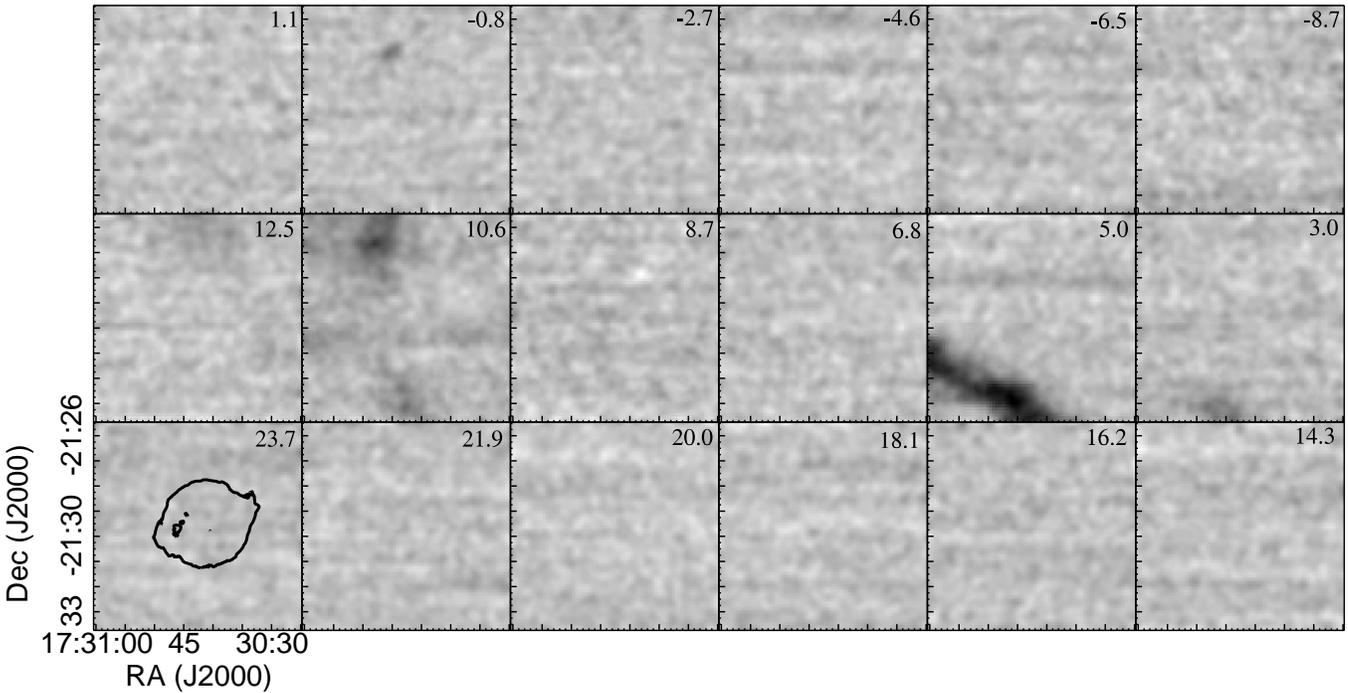}
\caption{Channel maps of $T_{\rm A}^*$ across the $^{12}$CO ($J$=2--1)
  transition ranging from $-10$ to 24.6\, km\,s$^{-1}$ with velocity
  width, $\Delta v = 1.9$\,km\,s$^{-1}$. The images have been smoothed
  with a 21-arcsec gaussian. The central velocity is labeled on each
  grid and the negative greyscale ranges from $-0.41$ to 1.2\,{\sc
    k}\,km\,s$^{-1}$.  Three clouds are detected: to the south of the
  remnant at velocities 3--5\,km\,s$^{-1}$, to the north at velocities
  9--11\,km\,s$^{-1}$ and a small cloud in the north at velocities
  0--2\,km\,s$^{-1}$.}
\label{fig:cochan} 
\end{figure*}
\begin{figure*}
\begin{center}
\includegraphics[angle=-90,width=18.3cm]{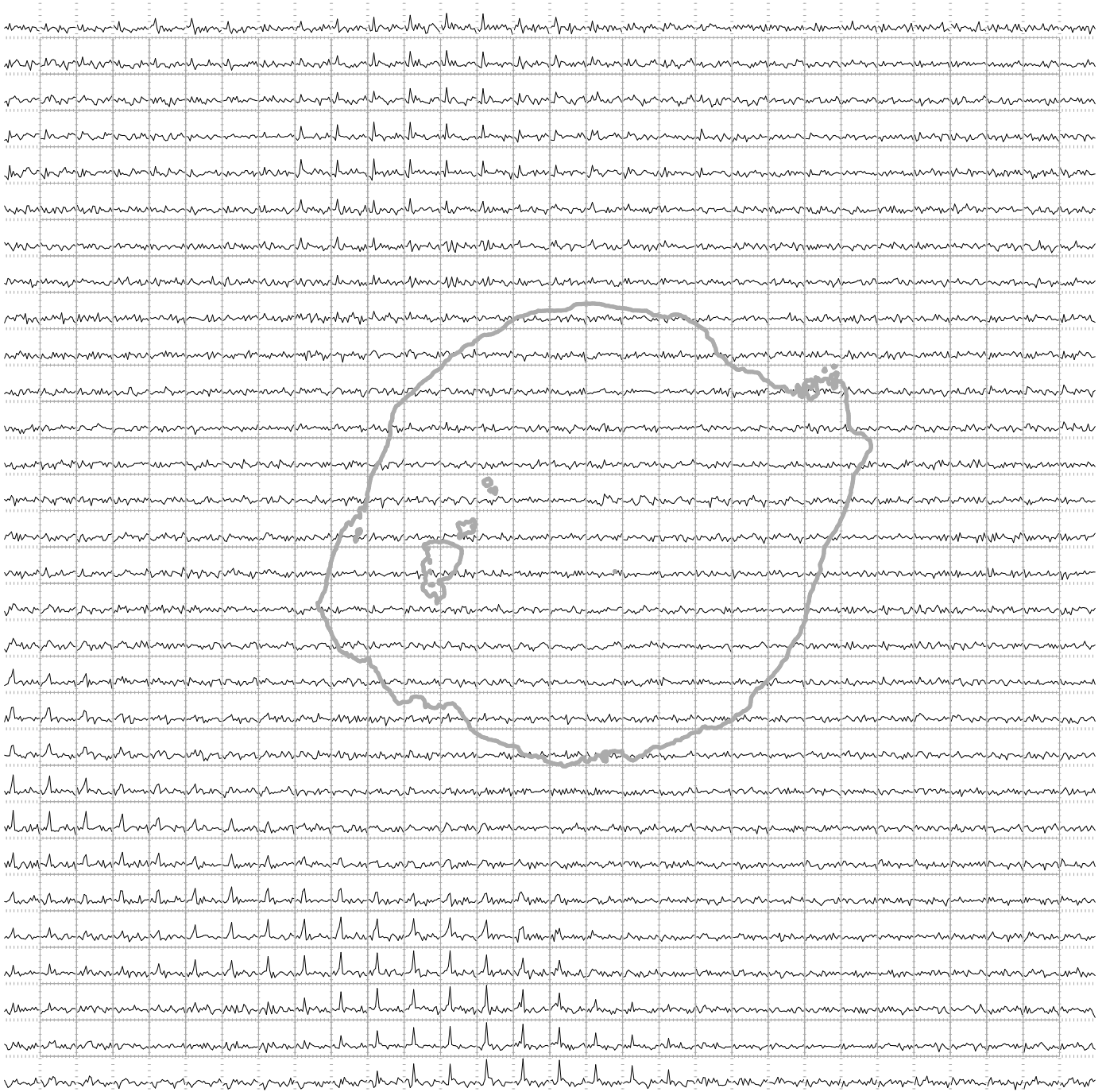}
\end{center}
\caption{Grid of $^{12}$CO ($J$=2--1) spectra towards Kepler's SNR
  from $-5 < v < +15$\,km\,s$^{-1}$ across the entire area. The black
  contour defines the shockfront as estimated from radio observations
  (Fig.~\ref{fig:clumps}).  The faint, narrow-line emission from the two
  molecular clouds in the south and in the north are clearly visible
  outside the area encompassed by the supernova remnant.}
\label{fig:lines} 
\end{figure*}
\begin{table*}
\begin{center}
\begin{tabular}{ccccccrr}\hline 
\multicolumn{1}{c}{Line} & \multicolumn{1}{c}{R.A.} & \multicolumn{1}{c}{Dec.}& \multicolumn{1}{c}{$t$ (mins.)} & \multicolumn{1}{c}{Date (2007)}  & \multicolumn{1}{c}{Beam ({\sc hpbw})} & \multicolumn{1}{c}{Spectral resolution} &\multicolumn{1}{c}{Pos. switch to:}\\ \hline
$^{12}$CO ($J$=2--1) &  17:30:42.0 & -21:29:35.4 & 1320 & March 08 - April 08 & 20.8$^{\prime \prime}$ & 977 kHz, 1.27\,km\,s$^{-1}$  &  -426.0$^{\prime \prime}$, +72.0$^{\prime \prime}$ \\
$^{12}$CO ($J$=2--1)  &  17:30:45.8 & -21:33:10.9 & 57&June 04 & 20.8$^{\prime \prime}$ &31 kHz, 0.08 km\,s$^{-1}$ &  +163.8$^{\prime \prime}$, +374.1$^{\prime \prime}$\\ 
$^{12}$CO ($J$=2--1)  &  17:30:46.0 & -21:27:23.7 & 167 &June 04 & 20.8$^{\prime \prime}$ & 31 kHz, 0.08 km\,s$^{-1}$ & +161$^{\prime \prime}$, +721.3$^{\prime \prime}$\\
$^{13}$CO ($J$=2--1)  &  17:30:45.8 & -21:33:10.9 & 167& June 04, 08 & 21.3$^{\prime \prime}$ & 31 kHz, 0.08 km\,s$^{-1}$ &  +163.8$^{\prime \prime}$, +374.1$^{\prime \prime}$\\ \hline
\end{tabular}
\end{center}
\caption{Summary of the CO observations used in this work.}
\label{tab:obs}
\end{table*}

We detect faint, narrow emission corresponding to the radial
velocities of the CO and H\,{\sc i} emission in the nearby $\rho$
Ophiuchus cloud complex between $-8 < v < 20$\,km\,s$^{-1}$ (de Geus
1992). Channel maps showing the total CO brightness temperature over
the range $-7.6 < v< 20.3$\,km\,s$^{-1}$ with velocity intervals of
1.9\,km\,s$^{-1}$ are shown in Fig.~\ref{fig:cochan}.  This range
encompasses all of the CO lines detected over the entire velocity
coverage.  Lines are observed at velocities of $-$1.4, 3.8 and
11.5\,km\,s$^{-1}$ with linewidths ({\sc fwhm}) $\Delta v \le
0.8$\,km\,s$^{-1}$, indicating cold, faint clouds. These lines are
narrower than expected for giant molecular clouds, typical
interstellar material (GMCs; Solomon, Sanders \& Scoville 1979,
$\Delta v \sim $4-7\,km\,s$^{-1}$), so-called dark clouds and
high-Galactic-latitude clouds (Lada et al.\ 2003).  The
high-resolution spectra at locations listed in Table~\ref{tab:obs}
were fitted with Gaussian profiles using the SPLAT--VO package (Draper
\& Taylor 2006) with CO intensities ranging from 0.5-1.4\,{\sc
  k}\,km\,s$^{-1}$.  It is difficult to ascertain distances to these
structures since the Galactic rotation models break down towards the
Galactic Centre.  Without an estimate of distance to the CO
structures, the observed size-velocity dispersion relationship for
molecular clouds in the Milky Way (Tsuboi \& Miyazaki 1998; Oka et
al.\ 1998) suggests that the CO clouds are smaller than 1\,pc across.

\begin{figure}
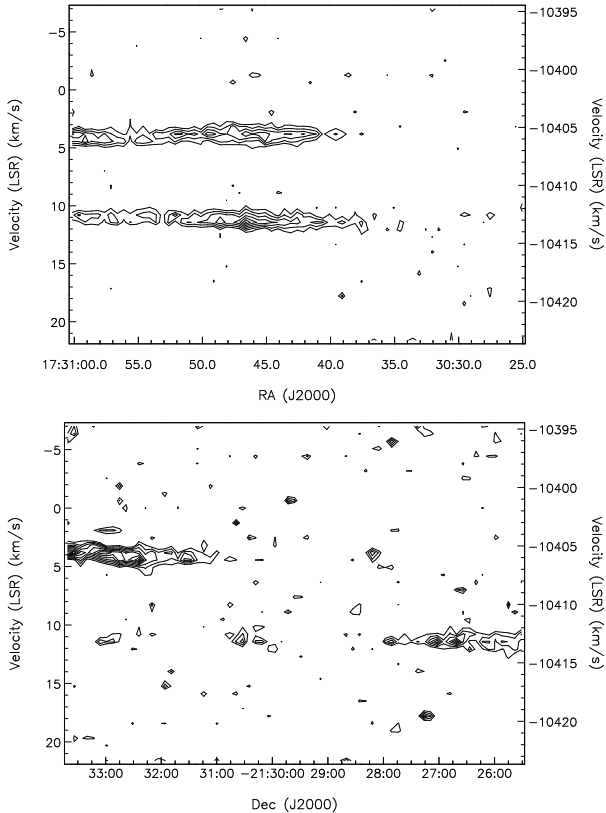

\begin{center}
{\subfigure{\includegraphics[angle=-90,width=8cm]{Gomez_fig6.ps}}}
{\subfigure{\includegraphics[angle=-90,width=8cm]{Gomez_fig7.ps}}}
\caption{High-resolution $^{12}$CO ($J$=2--1) spatial-velocity maps of the
region surrounding Kepler's SNR. {\em Top:} R.A.-velocity map with
contours starting at 0.04\,{\sc k} and incrementing by 0.03\,{\sc k}.
{\em Bottom:} Dec.-velocity map with contours starting 0.07\,{\sc k}
and incrementing by 0.03\,{\sc k}.}
\label{fig:covelpos} 
\end{center}
\end{figure}

We see no evidence for shocked molecular gas which would broaden the
lines to average linewidths $\Delta v$$>$15\,km\,s$^{-1}$, as seen in
the SNRs W28, W44 and W51C (De Noyer 1979; Junkes et al.\ 1992; Koo \&
Moon 1997; Seta et al.\ 1998; Reach, Rho \& Jarrett 2005). If an
interaction is present, we would expect to observe CO emission peaking
at the location of the shock front (e.g.\ Wilner et al.\ 1998) and a
velocity jump across the outer shock front.  Fig.~\ref{fig:lines}
shows the grid of spectra across the entire map, smoothed to a
velocity resolution of 3\,km\,s$^{-1}$ and restricted to $-5 < v <
+15\rm \,km\,s^{-1}$ for clarity.  The location of the supernova
shockfront (defined by the outer radio contour) is overlaid. The
molecular clouds detected here lie outside the shock front and
line-broadening is not observed in the profiles.  We also investigated
the velocity gradients across the clouds by integrating in R.A.\ and
declination (Fig.~\ref{fig:covelpos}). The CO clouds at 4 and
11\,km\,s$^{-1}$ line up at a common R.A., spanning 280\,arcsec with a
velocity range of $\sim$2\,km\,s$^{-1}$.  The clouds span a similar
range in Dec.\ and velocity in the RA. velocity-position map; the lack
of velocity gradient in these images confirms there is no interaction.

\subsubsection{$^{13}$CO($J$=2--1) observations}
\label{sec:co2}

\begin{figure}
\centering
\includegraphics[width=8.6cm]{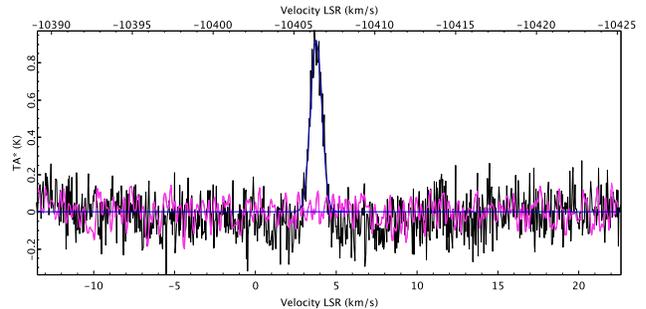}\hfill
\caption{ $^{12}$CO($J$=2--1) high-resolution spectra (black) observed
  at the location listed in Table~\ref{tab:obs} over the velocity
  range $-14 <v< 23$\,km\,s$^{-1}$.  The corresponding
  $^{13}$CO($J$=2--1) spectra are shown in purple.  The peak CO
  emission is $<$1 with velocity widths $\Delta v
  <$1\,km\,s$^{-1}$.}
\label{fig:cohighres} 
\end{figure}

We observed the brightest $^{12}$CO($J$=2--1) cloud, at
11\,km\,s$^{-1}$, in $^{13}$CO ($J$=2--1) and detected no emission;
the high-resolution spectrum is shown in Fig.~\ref{fig:cohighres}.
The 3\,$\sigma$ upper limit can be estimated using $I_{\rm CO} < 3
\sigma T_{\rm MB} \sqrt{\Delta v \, \delta v}$, where $\delta v$ is
the velocity width of the channel and $\Delta v$ is the velocity width
of the expected line. We find an upper limit of $I(^{13}{\rm CO}) <
0.08$\,{\sc k}\,km\,s$^{-1}$, for a width of 1\,km\,s$^{-1}$ (the
maximum width detected at the same location in $^{12}$CO).  The lower
limit on the brightness ratio $R$ is therefore $I(^{12}{\rm
  CO})/I(^{13}{\rm CO})>$17.  This is larger than those usually found
in GMCs where $R \sim 5$ (Solomon et al.\ 1979).

The optical depth, $\tau^{13}_{\nu}$, is estimated using Eq.~\ref{eq:tau}:
\begin{equation}
\tau_{\nu} = - {\rm ln} \left[1 - {T^{13}_{\rm MB} \over{T_0}} \left({{1\over{e^{\left(T_0/T_{\rm ex}\right)}-1}}}  - {{1\over{e^{\left(T_0/2.7\right)}-1}}}\right)^{-1}\right]
\label{eq:tau}
\end{equation}

\noindent
where $T_0$=$h\,\nu/k$.  Without further knowledge of the excitation
temperature of the gas $T_{\rm ex}$, we assume a nominal value of
15\,{\sc k} (similar to the dust temperatures implied by the
450-/850-$\mu$m flux ratios, M03).  The upper limit on the optical
depth is $\tau_{\nu}^{13}< 8\times 10^{-3}$ for $T_{\rm ex}$=15\,{\sc
  k}, assuming that $T_{\rm ex}$ is the same for the $^{12}$CO and
$^{13}$CO molecules.  The non-detection of $^{13}$CO indicates that
the $^{12}$CO emission is optically thin and confirms that we are
observing faint emission from a diffuse molecular cloud.  Under the
assumption that the excitation temperature is the same for both
molecules and emission fills the beam, the 3-$\sigma$ upper limit on
the optical depth of the $^{12}$CO emission is $\tau_{\nu}^{12} < 0.15$
(Eq.~\ref{eq:tauco}).
\begin{equation}
{T_{\rm mb}(^{12}{\rm CO})_{\nu} \over{T_{\rm mb}(^{13}{\rm CO})_{\nu}}} = {1 - \exp{\left(-\tau^{12}_{\nu}\right)} \over{1- \exp\left({-\tau^{13}_{\nu}}\right)}}
\label{eq:tauco}
\end{equation} 

The beam-averaged total column density, $N$, can be estimated using
the standard assumptions of radiative transfer in local thermodynamic
equilibrium (LTE) for an optically thin gas (Thompson \& MacDonald
1999):

\begin{equation}
  N = {3 k\over{8\pi^3}} {\int T_{\rm MB} {\rm d}v\over{ \nu S\mu^2 g_{\rm I} g_{\rm K} }}Q(T_{\rm ex}){\rm exp}\left({E_u\over{kT_{\rm ex}}}\right),
\label{eq:co12}
\end{equation}

\noindent
where $\nu$ is the line frequency, $S$ is the line strength, $\mu$ is
the permanent electric dipole moment (in e.s.u), $g_{\rm I}$ and
$g_{\rm K}$ are the reduced nuclear spin degeneracy and K-level
degeneracy. $E_{\rm u}$ is the energy of the upper level and $Q$ is
the partition function of the molecule (with excitation temperature,
$T_{\rm ex}$).  Values for the constants were obtained from Rohlfs \&
Wilson (2000) and the Jet Propulsion Laboratory (JPL) database
(Pickett et al.\ 1998). To determine the $\rm H_2$ column density, we
assume a $\rm ^{12}CO$ to $\rm H_2$ abundance ratio of $\sim 8.5
\times 10^{-6}$ (appropriate for diffuse clouds e.g.\ Liszt 2007; for
comparison, the typical values quoted for dense clouds is $8 \times
10^{-5}$).  The peak column densities of the molecular gas $N(\rm
H_2)$ seen towards Kepler's remnant estimated from the $\rm ^{12}CO$ emission
are 3.3, 4.4 and 9.1$\times 10^{19}$\,cm$^{-2}$.

The upper limit on the column density from the $^{13}\rm CO$ emission
can be estimated using a similar approach to Eq.~\ref{eq:co12}, with
an assumed linewidth and excitation temperature for the $^{13}$CO line
set from the observed values for the $^{12}$CO emission (e.g. Thompson
\& Macdonald 2003). Assuming a conversion ratio between $^{13}$CO to
$\rm H_2$ of $\sim 2 \times 10^{-7}$ (e.g.\ Langer \& Penzias 1990),
we estimate the $3\,\sigma$ limit on the molecular column density
$N({\rm H}_2) <$$10^{20}$\,cm$^{-2}$ at the location where we measure
both $^{12}$CO and $^{13}$CO (i.e. the peak emission in
$^{12}\rm CO$).\\

In order to confirm the validity of our LTE analysis we also modeled
the $^{12}$CO line emission with a non-LTE radiative transfer code
(van der Tak et al.\ 2007). The code assumes an isothermal, homogenous
medium which fills the telescope beam and solves the radiative
transfer equation using the escape probability formulation. We
modeled the integrated intensity of the $^{12}$CO emission for each
of the three positions given in Table~\ref{tab:obs}, assuming a
kinetic temperature $T_{\rm k}$=15\,{\sc k} and a range of different
H$_{2}$ densities. The results of our modelling fully confirm our LTE
analysis, confirming that the $^{12}$CO emission is both optically thin
and in LTE. Column densities derived by both methods agree within a
factor of 3. The radiative transfer models also suggest an upper limit
for the H$_{2}$ column density of $\sim$10$^{20}$\,cm$^{-2}$.

\subsubsection{H\,{\sc i} observations: searching for atomic structures}
\label{sec:hi}

To search for atomic foreground features towards Kepler, we have
revisited the H\,{\sc i} observations reported in RG99, made between
1996 and 1997, with the VLA in its CnB, C and D configurations. The
number of spectral channels was 127, centred at 0\,km\,s$^{-1}$ (LSR)
with a resolution of 1.3\,km\,s$^{-1}$. We studied a region of about
25\,arcmin $\times$ 25\,arcmin centred on Kepler's SNR.  The raw data
were downloaded from the VLA public archive and calibrated with the
{\sc aips} package, following standard procedures. The continuum was
subtracted in the visibility plane (recall Fig.~\ref{fig:clumps}) and
the Fourier transform and cleaning of the images was carried out using
the {\sc miriad} data processing package (Sault, Teuben \& Wright
1995). A first data cube was constructed using natural weighting. The
resolution and sensitivity achieved were slightly improved with
respect to those reported in RG99 but the sidelobe level was
unacceptable, so a second cube was constructed using uniform weighting
which drastically reduced the sidelobes and improved the resolution
but degraded the sensitivity.

We constructed an opacity ($\tau$) cube, defined as:
\begin{equation}
\tau = - {\rm ln }\left( 1 + T_{\rm B}(v)\over T_{\rm c}\right)
\label{eq:tau2}
\end{equation}

\noindent where $T_{\rm B}(v)$ is the brightness temperature at
velocity $v$ and $T_{\rm c}$ is the continuum emission.  To construct
the continuum image (Fig.~\ref{fig:clumps}), the line-free channels
used to determine the spectral baseline were combined in the {{\it u,v}} plane
and inverted using uniform weighting. The resulting beam is
9.2\,arcsec $\times$ 5.5\,arcsec, PA$=$51.6$^{\circ}$, and
sensitivity, 0.6\,mJy\,beam$^{-1}$. As originally noted in RG99, the
H\,{\sc i} emission is uniform in structure, this is confirmed in the opacity
cubes and we find no evidence for a correlation of H\,{\sc i}
structures with the submm emission. Several trials were made to
improve the sensitivity by Hanning smoothing the spectral profiles and
convolving the images to larger beams, with little success.  Addition
of single dish data from the LAB survey (Hartmann \& Burton 1997;
Kalberla et al.\ 2005) to the images did not add evidence of a
correlation.  

Using the VLA H\,{\sc i} and single-dish data we can estimate the column density of atomic gas at the location of the submm clumps:
\begin{equation}
N= 1.82 \times 10^{18}\,\int T {\rm d}v
\label{eq:hi}
\end{equation}
\noindent
where $v$ is the velocity (in km\,s$^{-1}$).  Integrating over the
velocity range in which large-scale H\,{\sc i} features are seen
($-10<v<22$\,km\,s$^{-1}$), we detect small variations in emission on
the H\,{\sc i} map, with minimum and maximum column densities of
0.6-0.9$\times 10^{21}$\,cm$^{-2}$.  The atomic gas density is at
least an order of magnitude higher than the molecular density
estimated from the CO.  The chopping procedure employed in the submm
observations of Kepler's SNR can not produce the submm structures from
chopping on-off these small variations, indeed chopping would have
removed any emission associated with the H\,{\sc i} as was seen in Cas
A (Wilson \& Bartla 2005).


\section{Comparing the Gas and Dust Towards Kepler}
\label{sec:dustmass}
\begin{figure*}
\includegraphics[width=18cm]{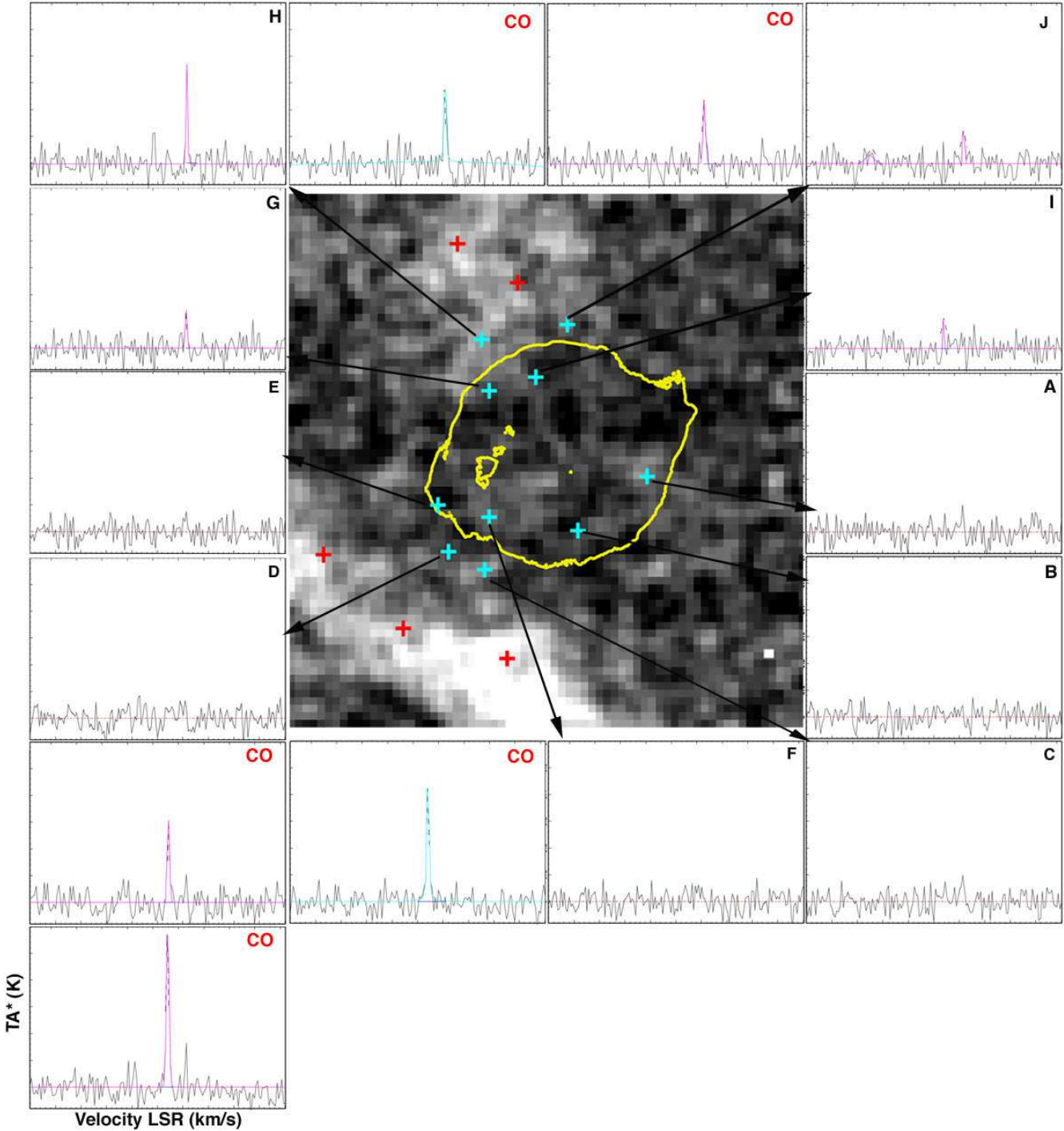}
\caption{Integrated $^{12}$CO($J$=2--1) emission towards Kepler's SNR
    over the velocity range $-$200--200\,km\,s$^{-1}$ smoothed by a
    3\,km\,s$^{-1}$ Gaussian. The crosses indicate regions where
    spectra were taken across the CO clouds and the remnant.  Cyan
    crosses show the locations of the submm clumps A--F, red crosses
    are positions chosen with strong CO emission for comparison. The integrated
    spectra from apertures at the locations of the submm clumps and at
    the CO clouds are shown alongside over the velocity range $-50 <
    v< 50$\,km\,s$^{-1}$ (unsmoothed). The greyscale ranges
    from $-$1.4 to 1.9\,{\sc k}\,km\,s$^{-1}$.}
\label{fig:mosaic} 
\end{figure*}
\begin{table*}
\begin{center}
\begin{tabular}{crrcccccccc}\hline 
 \multicolumn{1}{c}{Name} & \multicolumn{2}{c}{Coords}& \multicolumn{1}{c}{$T_{\rm peak}$} & \multicolumn{1}{c}{$v_{\rm c}$} & \multicolumn{1}{c}{\sc fwhm} & \multicolumn{1}{c}{r.m.s.} & \multicolumn{1}{c}{$I({\rm CO})$} & \multicolumn{1}{c}{$N(^{12}{\rm CO})$}& \multicolumn{1}{c}{$N({\rm H}_2)$} & \multicolumn{1}{c}{$S_{\rm 850}$}\\ 
\multicolumn{1}{c}{} &\multicolumn{1}{c}{$\Delta$ R.A.}&\multicolumn{1}{c}{$\Delta$ Dec.}&\multicolumn{1}{c}{\sc k}& \multicolumn{1}{c}{(km\,s$^{-1}$)}&\multicolumn{1}{c}{(km\,s$^{-1}$)}  & \multicolumn{1}{c}{\sc k} &\multicolumn{1}{c}{({\sc k}\,km\,s$^{-1}$)}& \multicolumn{1}{c}{($10^{14}$\,cm$^{-2}$)} & \multicolumn{1}{c}{($10^{19}$\,cm$^{-2}$)} & \multicolumn{1}{c}{(mJy)}\\ \hline 
CO S &+221.9&+95.9&  0.63 & 4.06 & 1.14 &0.05 & 1.12 & 6.0&7.1 & ..\\
&&&  0.51 & 11.17 & 0.46 & 0.05& 0.25 & 1.1&1.3&..\\
CO S & +146.3&+82.7& 0.67 & 4.05 & 1.10 & 0.05 &0.78 & 4.3&4.9&..\\ 
CO S & +51.1&+192.5& 0.46 & 4.30 & 1.00 & 0.06&0.49 &2.7&3.1&..\\
CO N & +96.9 &$-$190.9&0.34 & 11.48 & 1.17 &0.05& 0.42  &2.2&2.7&..\\
CO N & +47.6&$-$153.5 &1.14 & 11.12 & 0.18 &0.06 &0.81 &4.4&5.1&..\\
A &$-$83.3&+23.3& 0.14& 11.4 & 0.96 & 0.05 & 0.16   &8.6&10.2&28.8 $\pm$ 3.5\\
B & $-$17.5&+71.8&.. & .. &.. & 0.05 &$\le$0.15 &$\le$0.8 &$\le$1.0& 23.0 $\pm$ 2.9 \\
C &+69.3 &+110.3&0.13 & 11.21 & 1.59 & 0.05& 0.22& 1.2 &1.5& 41.5 $\pm$ 4.6\\
D &+101.5&+93.3& .. & ..&.. & 0.05& $\le$ 0.15 &$\le$0.8 &$\le$1.0 & 45.8 $\pm$ 4.9\\
E &+109.9 &+47.3&.. & .. & ..& 0.05 &$\le$0.15&$\le$0.8&$\le$1.0 &62.1 $\pm$ 6.6 \\
F &+94.5 &+58.3&.. & .. &.. & 0.06 &$\le$0.18&$\le$1.0 &$\le$1.1&21.8 $\pm$ 2.7\\
G & +65.1 &$-$57.0 & 0.20 & 11.27 & 0.87& 0.05 & 0.19 & 10.2&1.2& 30.6 $\pm$ 3.6\\
H &+70.7&$-$104.7& 0.55 & $-$1.59 & 0.48 & 0.05& 0.28 & 1.6&1.8&63.8 $\pm$ 6.7\\
  &&& 0.55 & 11.36 & 0.76 & 0.05&0.42 & 2.2&2.7& ..\\ 
I & +21.7 &$-$71.7&0.23 & 11.71 & 0.68 & 0.05& 0.20 &1.1 &1.2& 31.4 $\pm$ 3.3\\
J & $-$7.7&$-$118.7& 0.27 & 11.70&0.72 &0.05 &0.21&1.1&1.3& 33.5 $\pm$ 3.8\\ \hline
\end{tabular}
\end{center}
\caption{Properties of the detected $^{12}$CO ($J$=2--1) line
  emission towards Kepler's SNR.  Columns show (1) name (2) offset
  coordinates from centre in arcsec ($\rm RA=17^h 30^m41^s.3$ and
  $\rm Dec. = -21^{\circ} 29^{\prime}29^{\prime \prime}$ where
  postive R.A.\ and Dec.\ indicates a move to the East and South);
  (3) peak $T_{\rm MB}$ (4) central velocity  (5) velocity width
  ({\sc fwhm}) (6) integrated intensity (7) rms (8) and (9) column densities of CO and H$_2$
  gas estimated using Eq.~\ref{eq:co12} and (10) the submm flux measured in the apertures A--J
  after synchrotron subtraction.  Errors are estimated using the procedure outlined in Dunne \& Eales (2001). }
\label{tab:coprops}
\end{table*}

The high resolution $^{12}$CO ($J$=2--1) map integrated over the
velocity range $-200 < v< 200$\,km\,s$^{-1}$, is shown in
Fig.~\ref{fig:mosaic} with the outer contour of the radio image
overlaid.  The map has been smoothed with a 21-arcsec Gaussian. Cyan
crosses indicate the submm dust clumps, A--J; red crosses indicate CO
peaks. Spectra at each of these locations were extracted; where CO was
detected, the lines were fitted with Gaussian profiles with the
derived properties given in Table~\ref{tab:coprops}.  Also listed in
Table~\ref{tab:coprops} are the fluxes in the submm clumps, A--J,
determined by placing an aperture of radius 23\,arcsec on the
synchrotron-subtracted 850-$\mu$m image. The sum of the fluxes in
these apertures is 0.38\,Jy (cf.\ 0.58\,Jy in M03).  The submm clumps
outside the remnant are located on average, in regions of higher CO
emission than those inside the remnant, however a significant fraction
of the submm flux comes from clumps where there is no, or little CO
emission.  Comparing the distribution of the CO and submm emission
{\it spatially}, we can see that molecular clouds are observed at the
location of submm clumps G and H (combined flux of 94\,mJy) and
fainter emission at clumps A, C, I and J (combined submm flux
135\,mJy).  We do not detect CO emission at the locations of the submm
clumps B, D, E and F (combined submm flux 153\,mJy). 

The CO column densities towards these clumps is a better indicator of
the {\it quantity} of line-of-sight emission. At apertures A--J, these
range from $<$10$^{14}$--10$^{15}$\,cm$^{-2}$ with molecular gas
densities $10^{19}$--$10^{20}$\,cm$^{-2}$.  The gas column densities
estimated from the submm fluxes range from $10^{21}$ to
$10^{22}$\,cm$^{-2}$ (with gas-to-dust ratio of 160); as confirmed by
the column densities estimated from the (lower-resolution) 100$\mu$m
{\it IRIS} maps (Miville-Desch\^{e}nes \& Lagache 2005). This a factor
of 10-100 times higher than that estimated from the CO data.  If the
submm clumps were related to the same molecular gas as traced by the
foreground CO emission, the conversion ratio $N\rm (H_2)/CO$ in these
clumps would have to be less than $10^{-7}$ which is far lower than
the values ($>$$10^{-5}$) expected for clouds with densities of
$10^{21}$--$10^{22}$\,cm$^{-2}$ (e.g. Listz 2007).  Conversely, using
the appropriate conversion factor for the molecular gas densities
estimated from the submm clumps, we would expect to observe CO column
densities $N(\rm CO)>$$10^{16}$--$10^{17}$\,cm$^{-2}$ at the location
of the submm emission.  This suggests that the molecular gas
contribution to the submm emission towards Kepler is negligible, even
where CO is co-spatial with the submm clumps on the sky.

Are there any physical effects that could result in our underestimating
the column density of the gas? In dense molecular cores, CO and other
molecules may become depleted from the gas phase by freezing out onto
the surfaces of dust grains (e.g. Redman et al.\ 2002). The timescale
for molecular freezeout is tightly correlated to the gas density and
only becomes significant for H$_{2}$ densities $>$10$^{5}$ cm$^{-3}$
(Caselli et al.\ 1999). If this were the case then we would expect to
see evidence for higher gas densities in the form of higher optical
depths and the presence of rarer isotopologues of CO. The properties
of the $^{12}$CO emission that we have detected are strongly
suggestive of diffuse clouds in which molecular depletion does not
play a significant role in controlling the abundance of CO.

In diffuse molecular clouds, the CO abundance depends on the overall
H$_{2}$ column density (Liszt et al 2007), likely due to the increased
UV photodissociation rates for CO in low density gas (van Dishoeck \&
Black 1988). However, in order to match the column densities indicated
by the submm emission, the density of the gas would increase to a
point where photodissociation is largely unimportant except at the
exterior of the cloud. The CO to H$_{2}$ abundance ratio for gas with
an H$_{2}$ column density of 10$^{21}$--10$^{22}$ is close to that
expected for a dark molecular cloud and not
consistent with the column densities of CO that we observe. We thus
conclude that neither depletion or photodissociation are likely to
have a significant effect upon our derived column densities.

\subsection{The revised dust mass in Kepler's remnant}
\label{sec:dust}

\begin{figure}
\centering
{\includegraphics[width=6.2cm,height=8.5cm,angle=-90]{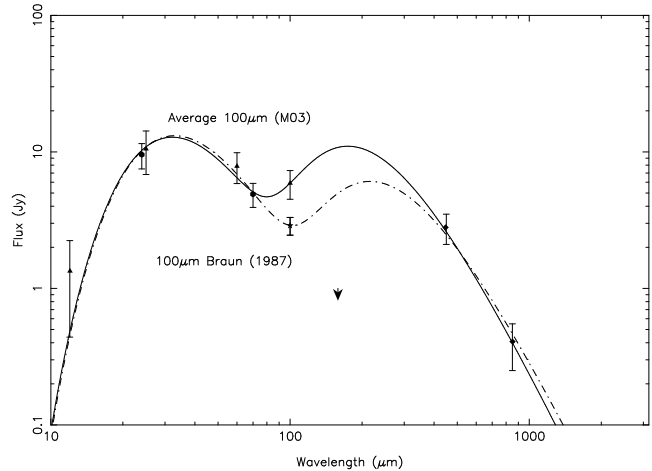}}\hfill
\caption{IR--submm SED for Kepler's SNR, comparing the best fits with
  two different 100-$\mu$m fluxes, including {\it Spitzer} data and
  submm points (this work).  The solid line represents the best fit
  with $S_{100}\sim 5.9 \pm 1.4$\,Jy.  The dot-dashed curve represents
  the fit to the lower 100-$\mu$m flux, $2.9 \pm 0.9$\,Jy. The upper
  limit shown by the arrow is the total integrated flux obtained from
  {\it Spitzer} at 160\,$\mu$m (Blair et al.\ 2007). }
\label{fig:sed} 
\end{figure}

The IR--submm spectral energy distribution (SED) of Kepler's SNR,
using the synchrotron-subtracted submm fluxes is shown in
Fig.~\ref{fig:sed}.   We fit a two-component modified
blackbody to the SED which is the sum of two modified Planck
functions, each with a characteristic temperature, $T_w$ and $T_c$
(Eq.~\ref{eq:sed}):
\begin{equation}
S_{\nu} = N_w \times \nu^{\beta}B(\nu, T_w) + N_c\times \nu^{\beta}B(\nu, T_c)
\label{eq:sed}  
\end{equation}

where $N_w$ and $N_c$ represent the relative masses in the warm and
cold component, $B(\nu,T)$ is the Planck function and $\beta$ is the
dust emissivity index.  The model was fitted to the SED (constrained
by the 12-$\mu$m flux) and the resulting parameters ($N_w$, $N_c$,
$T_w$, $T_c$, $\beta$) which gave the minimal $\chi^2$ were found.  
  
The solid curve shows the fit to the SED with $S_{100} = 5.6 \pm
1.4$\,Jy (the average from Arendt 1989 and Saken, Fesen \& Shull 1992)
whilst the dot-dashed curve fits the SED with the lower 100-$\mu$m
flux of 2.9\,Jy (estimated from pointed observations, Braun 1987). The
higher 100-$\mu$m flux predicts the {\it lower} dust mass for the SED
model.  Both fits include the recent {\it Spitzer} fluxes at 24 and
70\,$\mu$m (Blair et al.\ 2007) but the 160\,$\mu$m upper limit is not
included in the fit due to uncertainties with the background at this
wavelength.  Neither fit to the IR data alone would provide evidence
for a cold dust component, highlighting the importance of accurate
submm fluxes when estimating the dust mass. Note that
the SED is equally well fit in both cases and to distinguish between
the two, more accurate fluxes are needed around the peak of the cold
emission at 200--400\,$\mu$m.  

In addition to using the range of {\it IRAS} 100-$\mu$m fluxes in the
literature, we also applied a bootstrap analysis to our SED fitting to
determine the errors on the model. The photometry measurements were
perturbed according to the errors at each wavelength (using the higher
100-$\mu$m flux which produces the lowest mass). The new fluxes were
fitted in the same way, the SED parameters and dust masses recorded,
with the procedure repeated 3,000 times.  The median results along
with errors (estimated from the 68 per cent confidence intervals) are
listed in Table~\ref{tab:dustmodel}.

The dust mass is calculated from the SED model using
Eq.~\ref{eq:sed2}.
\begin{equation}
M_d = {S_{850} D^2 \over{\kappa_{850}}}\times \left[ {N_w \over{B({\rm 850}~\mu{\rm m}, T_w)}} +{ N_c\over{B({\rm 850}~\mu{\rm m}, T_c)}}\right]
\label{eq:sed2}  
\end{equation}
where $\kappa_{850}$ is the dust absorption coefficient and the
distance, $D$ is taken to be the lower published limit of 3.9\,kpc
(Sankrit et al. 2008). To determine the dust mass, we take two extreme
values for $\kappa$: (i) $\kappa_{850}= 0.07$\,m$^2$\,kg$^{-1}$,
typical quoted value for grains in the diffuse interstellar medium
(D03 and references therein) and (ii)
$\kappa_{850}=0.7$\,m$^2$\,kg$^{-1}$, as required for the submm
polarimetry observations of Cas~A (Dunne et al. 2009). This is similar
to the values ($\sim 0.4$\,m$^2$\,kg$^{-1}$) predicted by the
supernova dust model in Bianchi \& Schneider (2007).  The range of
dust masses obtained from the SED model (Eqs.~\ref{eq:sed} \&
\ref{eq:sed2}) is therefore 0.1--1.2\,$\rm M_{\odot}$ depending on
$\kappa$.  Note that $\kappa$ is the largest source of error when
estimating the mass of dust from IR/submm emission.

\begin{table}
\centering
  \begin{tabular}{ccccc} \hline
 \multicolumn{3}{c}{Bootstrap parameters} &\multicolumn{2}{c}{Dust mass (M$_{\odot}$)}\\ \hline
 \multicolumn{1}{c}{$\beta$}&\multicolumn{1}{c}{$T_{\rm w}$
      ({\sc k})}&\multicolumn{1}{c}{$T_{\rm c}$ ({\sc k})} &\multicolumn{1}{c}{$\kappa_1 \rm \,(m^2\,kg^{-1})$} &\multicolumn{1}{c}{$\kappa_2 \rm \,(m^2\,kg^{-1})$}  \\\hline 
$2.0 \pm 0.2$ & $88^{+6}_{-4}$ & $ 16\pm 1$ & $0.12^{+0.02}_{-0.03}$ & $1.2^{+0.18}_{-0.30}$  \\   \hline
\end{tabular}
\caption{The best-fit parameters for Kepler's SNR
  estimated from the bootstrap technique. The dust mass is
  the median of the distribution from the bootstrap technique with
  the error quoted from the 68 per cent confidence interval. $\kappa_1=0.7\rm \,m^2\,kg^{-1}$ (appropriate for SNR dust, Dunne et al.\ 2009) and $\kappa_2=0.07\rm \,m^2\,kg^{-1}$ (typical ISM values) which gives the maximum mass.  }
\label{tab:dustmodel}
\end{table}

\section{Conclusions} 
\label{sec:conc}

We have presented information on the reduction and analysis of SCUBA
submm data of Kepler's SNR including a more accurate subtraction of
the synchrotron component. The residual fluxes are slightly lower than
those previously reported.  We find that the ring-like structure seen
in the submm cannot be reproduced by chopping on-off large scale
structures.  Large-scale FIR/submm imaging is crucial to determine
whether the observed submm structures are unique to the SNR, or are
common in this region. Such observations will soon be possible with
SCUBA-2 and the {\it Herschel Space Observatory}.

We investigated whether foreground molecular or atomic structures could be
responsible for the submm emission concluding that:
\begin{itemize}
\item There are three molecular clouds in the vicinity of Kepler's
  SNR, located on the periphery of the remnant in the north and
  south-east and extending further out in the south.  The clouds are
  faint and cold ($T_{\rm A}^*$ $<$1\,{\sc k}, $\Delta
  v < 1$\,km\,s$^{-1}$).  Molecular gas column densities towards the
  remnant are estimated to be $\le$$10^{20}$\,cm$^{-2}$.
\item $^{13}$CO line emission is not detected at the locations of the
  peak $^{12}$CO emission.  The 3-$\sigma$ upper limit on the
  molecular column density is $<$$10^{20}$\,cm$^{-2}$ indicating
  that the clouds are diffuse and optically thin.  This is confirmed
  with an independent analysis.
\item The column densities estimated from the submm are 10--100 times
  higher than the column densities estimated from the molecular gas.
  This difference cannot be explained by reasonable variations in the
  conversion factor between CO and $\rm H_2$ emission nor can it be
  explained by depletion or photodissociation effects.
\end{itemize}
The molecular contribution to the submm emission towards Kepler's
remnant is therefore negligible.  The dust mass associated with the
remnant ranges from 0.1--1.2\,$\rm M_{\odot}$, depending on the
absorption coefficient.  This value is 100 times larger than
seen by {\it Spitzer} and concurs with the results for Cas~A (Dunne
et al.\ 2009) that supernovae are significant sources of dust.

\appendix
\label{sec:bruce}
\section{Spurious structure due to chopping?}

\begin{figure*}
\includegraphics[width=15.5cm,height=9.5cm]{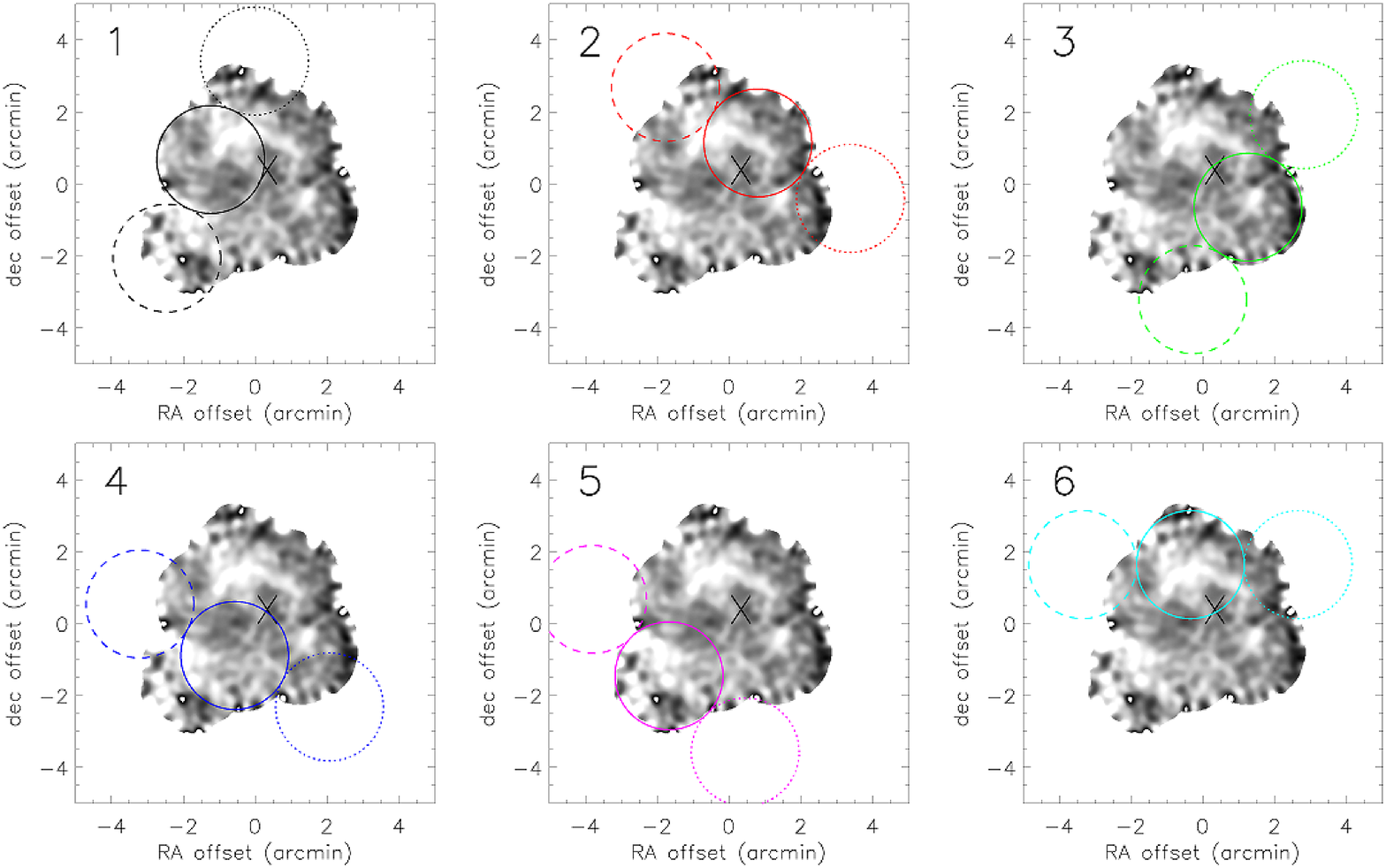}
\caption{The individual six SCUBA jiggle maps and corresponding chop
  positions.  The greyscale image is the total 850-$\mu$m signal-to-noise
  image.}
\label{fig:bruce}
\end{figure*}

\begin{figure*}
\includegraphics[width=17cm,height=10cm]{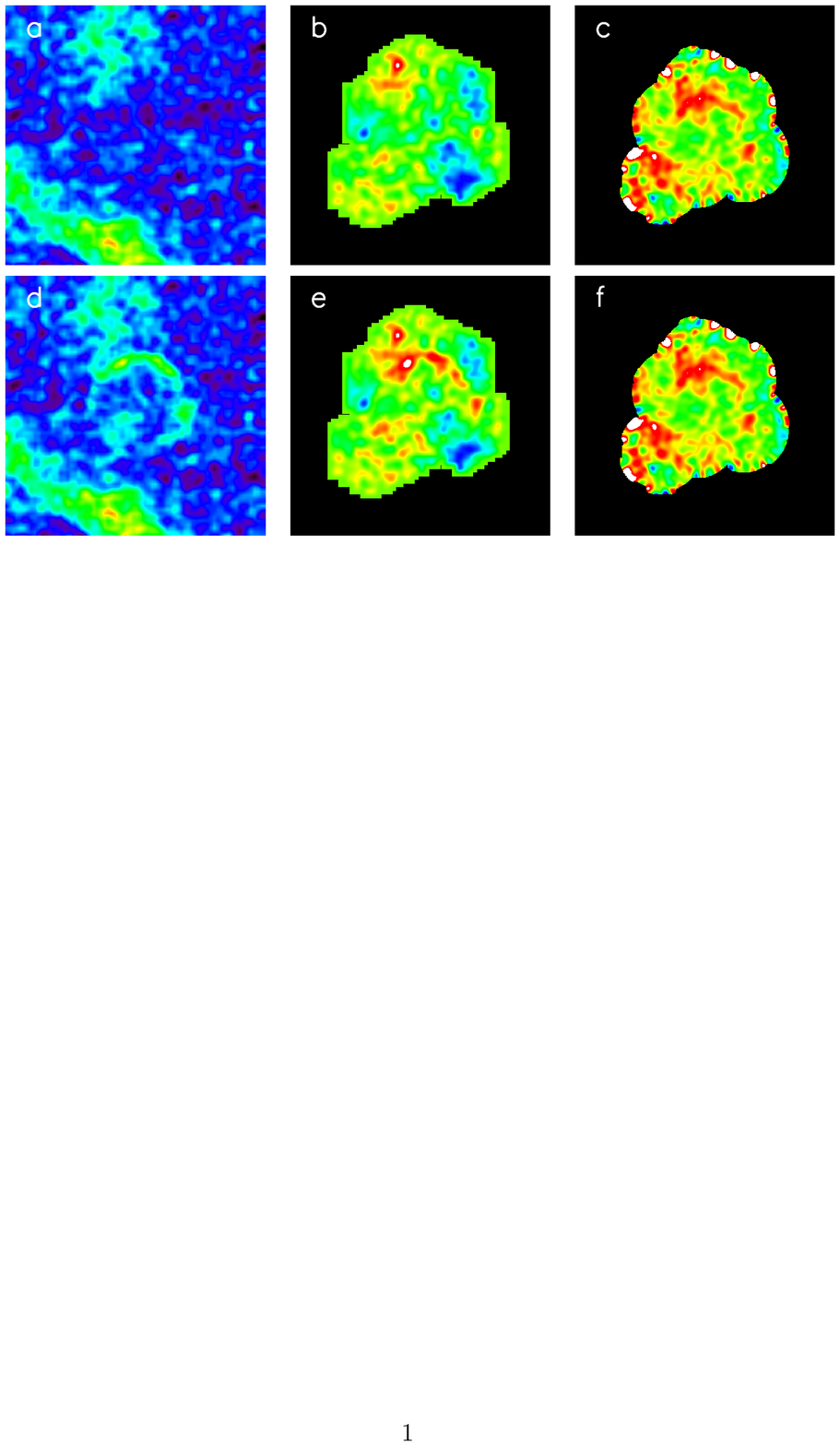}
\caption{Simulating the SCUBA maps: the input maps with foreground CO
  used as an example of large scale structures surrounding Kepler (a)
  combined with the radio SNR (d).  The results are given in b and
  e. Panels c and f show the original SCUBA (S/N) map at 850\,$\mu$m for
  reference.}
\label{fig:sims}
\end{figure*}

The observing technique used when obtaining the original SCUBA data
(\S\ref{sec:submm}) could affect the apparent distribution of submm
emission around Kepler's SNR if we had chopped onto structure
unrelated to the remnant. We modeled the effect of chopping onto
interstellar structures not associated with the remnant using the chop
positions and throws chosen for Kepler (Fig.~\ref{fig:bruce}).  The
simulations were performed by sampling the sky model at the
appropriate on-off source chop locations, white noise was added before
reconstructing the on-source flux measurements from the three
different chop/nod positions. We included different input models of
the sky, here we discuss the results using (i) large scale
interstellar submm structures (with the CO emission as a guide,
Fig.~\ref{fig:sims}a) and (ii) large scale interstellar submm
structures including a SNR ring-like structure at the location of
Kepler (based on the radio image, Fig.~\ref{fig:sims}d).


The results are shown in Figs.~\ref{fig:sims}b and \ref{fig:sims}e
along with the original SCUBA data in the same colour scale (c and
f). We see some spurious structure in the simulated output for case
(i) as a result of chopping on to the bright extended structure in the
south seen in the input sky. The effect of this chop is diminished
when averaged over the two off-source chop positions (jiggles three
and five).  The strong ring-like emission structure seen in the SCUBA
data (c) is not reproduced in the simulation, suggesting that the ring
cannot be a result of chopping on or off large scale background
structures.  This is supported by the output from case (ii) which
includes a SNR model in the input (e).  The ring-like structure survives
after chopping on-off the larger surrounding structures.  We conclude
that the emission in the SCUBA map is due to a ring-like source of
submm radiation and is not an artifact of the observing technique.

\section*{Acknowledgements}

H.L.G.\ would like to acknowledge the support of Las Cumbres
Observatory Global Telescope Network. E.M.R.\ is partially supported
by grants PIP-CONICET 114-200801-00428, UBACyT X482 and A023, and
ANPCYT-PICT-2007-00902. The data presented here were awarded under
JCMT allocations S07AU22 and M07AU23 and we thank all the JCMT staff
for their help with these programmes. This research has made use of
the NASA/IPAC Infrared Science Archive, which is operated by the Jet
Propulsion Laboratory, California Institute of Technology, under
contract with the National Aeronautics and Space Administration. We
gratefully thank William Blair, Thomas Dame, Edward Gomez and Oliver
Krause for comments and the referee for their insightful comments.


\begin{thebibliography}{}
\bibitem{} Arendt R.G., 1989, ApJS, 70, 189 
\bibitem{} Bianchi S., Schneider R., 2007, MNRAS, 378, 973
\bibitem{} Blair W.P., Ghavamian P., Long K.S., Williams B.J., Borkowski
           K.J., Reynolds S.P., Sankrit R., 2007, ApJ, 662, 998
\bibitem{} Borkowski K.J., Kazimierz J., Williams B.J., Reynolds S.P.,
           Blair W.P., Ghavamian P., Sankrit R., Hendrick S.P.,
           2006, ApJ, 642, L141  
\bibitem{} Braun R., 1987, A \& A, 171, 233 
\bibitem{} Caselli, P., Walmsley, C.~M., Tafalla, M., Dore, L., \& Myers, P.~C.\ 1999, ApJ, 523, L165
\bibitem{} Clayton D.D., Arnett D., Kane J., Meyer B.S., 1997, AJ, 486, 824
\bibitem{} Dame T., Hartmann D., Thaddeus P., 2001, ApJ, 547, 792 
\bibitem{} de Geus E., 1992, A \& A, 262, 258 
\bibitem{} DeLaney T., Koralesky B., Rudnick L., Dickel J.R.,
           2002, ApJ, 580, 914 
\bibitem{} DeNoyer L.K., 1979, ApJ, 232, L165 
\bibitem{} D\'{e}sert F.-X., Mac\'{\i}as-P\'{e}rez J.F., Mayet F.,
           Giardino G., Renault C., Aumont J., Beno\^{i}t A., Bernard J.-Ph,
           Ponthiey N., Tristram M., 2008, A \& A, 481, 411
\bibitem{} Draper P., Taylor M.B., 2006, A Spectral Analysis Tool,
           Starlink User Note 214
\bibitem{} Dunne L., Eales S., 2001, MNRAS, 327, 697
\bibitem{} Dunne L., Eales S., Ivison R.J., Morgan H., Edmunds M.,
           2003, Nature, 424, 285 {\bf [D03]}
         \bibitem{} Dunne L., Maddox S.J., Ivison R.J., Rudnick L.R.,
           DeLaney T.M., Matthews B.C., Crowe C.M., Gomez H.L., Eales
           S.A., Dye S., 2009, MNRAS, 320
\bibitem{} Dwek E., Galliano F., Jones A.P., 2007, ApJ, 662, 927
\bibitem{} Eales S., Lilly S., Webb T., Dunne L., Gear W.,
           Clements D., Yun M., 2000, AJ, 120, 2244 
\bibitem{} Eales S., Bertold F., Ivison R.J., Carilli C.L., Dunne
           L., Owen F., 2003, MNRAS, 344, 169 
\bibitem{} Gomez H.L., Eales S.A., Dunne L., 2007, IJAsB, 6, 159 
\bibitem{} Hartmann D., Burton W. B., 1997, Atlas of Galactic Neutral
           Hydrogen, Cambridge University Press, UK
\bibitem{} Holland W.S.\ et al., 1999, MNRAS, 303, 659 
\bibitem{} Isaak K., Priddey R.S., McMahon R.G., Omont A.,
           Peroux C., Sharp R.G., Witthington S., 2002, MNRAS, 329, 149 
\bibitem{} Jones A.P., Tielens A.G.G.M., Hollenbach D.J., McKee
           C.F., 1994, ApJ, 433, 797 
\bibitem{} Junkes N., Furst E., Reich W., 1992, A \& ASS, 96, 1 
\bibitem{} Koo B-C., Moon D-S., 1997, ApJ, 485, 263 
\bibitem{} Krause O., Birkmann S.M., Reike G., Lemke D., Klaas
           U., Hines D.C., Gordon K.D., 2004, Nature, 432, 596 
\bibitem{} Lada C.J., Bergin E.A., Alves Jo\~{a}o-F., Huard T.L., 2003, ApJ, 586, L286
\bibitem{} Langer W.D., Penzias A.A., 1990, ApJ, 357, 477
\bibitem{} Liszt H.S., 2007, A \& A, 476, 291
\bibitem{} Margulis M., Lada C.J., 1985, ApJ, 299, 925
\bibitem{} Meikle W.P.S., Mattila S., Pastorello A., Gerardy C.L.,
           Kotak R., Sollerman J., Van Dyk S.D., Farrah D.,
           2007, ApJ, 665, 608 
\bibitem{} Miville-Desch\^{e}nes M.-A., Lagache G., 2005, ApJSS, 157, 302 
\bibitem{} Morgan H.L., Edmunds M.G, 2003, MNRAS, 343, 427 
\bibitem{} Morgan H.L., Dunne L., Eales S., Ivison R.J., Edmunds
           M.G., 2003, ApJ, 597, L33 {\bf [M03]}
\bibitem{} Nozawa T., Kozasa T., Umeda H., Maeda K., Nomoto K.,
           2003, ApJ, 598, 785 
\bibitem{} Pickett H. M., Poynter R. L., Cohen E. A., Delitsky M. L.,
           Pearson J. C., Müller H. S. P., 1998, J.\ Quant.\ Spectrosc.\
           Radiat.\ Transfer, 60, 883
\bibitem{} Reach W.T., Rho J., Jarett T.H., 2005, ApJ, 618, 297 
\bibitem{} Redman, M.~P., Rawlings, J.~M.~C., Nutter, D.~J., Ward-Thompson, D., 
\& Williams, D.~A.\ 2002, MNRAS, 337, L17 
\bibitem{} Reynolds S., Borkowski K.J., Hwang U., Hughes J.P.,
           Badenes C., Laming J.M., Blondin J.M., 2007, ApJ, 668, L135
\bibitem{} Reynoso E.M., Goss W.M., 1999, AJ, 118, 926 {\bf [RG99]}
\bibitem{} Rho J., Kozasa T., Reach W.T., Smith J.D., Rudnick L.,
           DeLaney T., Ennis J.A., Gomez H., Tappe A., 2008, ApJ, 673, 271
\bibitem{} Rohlfs K., Wilson T. L., 2000, Tools of Radio Astronomy,
           3rd Edition, Springer-Verlag, Berlin
\bibitem{} Saken J.M., Fesen R.A., Shull J.M., 1992, ApJS, 81, 715 
\bibitem{} Sandell G., Jessop N., Jenness T., 2001, SCUBA Map Reduction
           Cookbook, Starlink Cookbook 11.2 
\bibitem{} Sankrit R., Blair W.P., Frattare L.M., Rudnick L., DeLaney T.,
           Harrus, I.M., Ennis J.A., 2008, AJ, 135, 538 
\bibitem{} Sault R.  J., Teuben P. J., Wright M. C. H., 1995, {\em in}
            Astronomical Data Analysis Software and Systems IV, eds
            Shaw R.A., Payne H.E., Hayes J.J.E., ASP, 77, 433, San Francisco
\bibitem{} Schneider R., Ferrara A., Salvaterra R., 2004, MNRAS, 351, 1379
\bibitem{} Seta M, Hasegawa T, Dame T.M., Sakamoto S., Ok T., Handa T.,
           Hayashi M., Morin J-I., Sora K., Usuda K.S., 1998, ApJ, 505, 286 
\bibitem{} Smail I., Ivison R.J., Blain A.W, 1997, ApJ, 490, L5
\bibitem{} Solomon P.M., Sanders D.B., Scoville N.Z., 1979, ApJ, 232 L89
\bibitem{} Sugerman, B.E.K., Ercolano B., Barlow M.J., Tielens A.G.G.M.,
           Clayton G.C., Zijlstra, A.A., Meixner, M., Speck, A.,
           2006, Science, 313,  196
\bibitem{} Thompson M.A., MacDonald G.H., 1999, A \& AS, 135, 531
\bibitem{} Thompson M.A., MacDonald G.H., 2003, A \& AS, 407, 237
\bibitem{} Todini P., Ferrara A., 2001, MNRAS, 325, 726
\bibitem{} Travaglio C., Gallino R., Amari S., Zinner E., Woosley S.,
           Lewis R.S., 1999, ApJ, 510, 325
\bibitem{} van der Tak, F.~F.~S., Black, J.~H., Sch{\"o}ier, F.~L., Jansen, D.~J., \& van Dishoeck, E.~F.\ 2007, A \& A, 468, 627
\bibitem{} van Dishoeck E.F., Black J.H., 1988, ApJ, 334, 771
\bibitem{} Williams B.J.\ et al., 2006, ApJ, 652, L33 
\bibitem{} Wilner D.J., Reynolds S.P., Moffett D.A., 1998, AJ, 115, 247 
\bibitem{} Wilson T.L., Batrla W., 2005, A \& A, 430, 561

\end{thebibliography}
\end{document}